\documentclass[12pt]{article}
\usepackage{graphicx}
\usepackage[margin=1.25in]{geometry}

\usepackage{xspace}
\usepackage{wrapfig}
\usepackage{amsmath}

\newcommand\pubnumber{CIPANP2015-Sandacz}
\newcommand\pubdate{\today}

\def\ncbj{Division of High Energy Physics\\
National Centre for Nuclear Research, PL 00-681 Warsaw, Poland}
\def\support{\footnote{Work supported by the Polish NCN Grant 
          DEC-2011/01/M/ST2/02350.}}

\def\Title#1{\begin{center} {\Large #1 } \end{center}}
\def\Author#1{\begin{center}{ \sc #1} \end{center}}
\def\Address#1{\begin{center}{ \it #1} \end{center}}

\newcommand{\aphi}{$\phi$\xspace}
\newcommand{\aphis}{$\phi_{s}$\xspace}

\newcommand{\AUTphiS} {$A^{\sin\phi_S}_{\text{UT}}$}

\newcommand{\AUTphi} {$A^{\sin(2\phi-\phi_S)}_{\text{UT}}$}
\newcommand\pubblock{\rightline{\begin{tabular}{l} \pubnumber\\
         \pubdate  \end{tabular}}}
\newenvironment{Abstract}{\begin{quotation}  }{\end{quotation}}
\newenvironment{Presented}{\begin{quotation} \begin{center} 
             Presented at\end{center}\bigskip 
      \begin{center}\begin{large}}{\end{large}\end{center} \end{quotation}}

\let\OLDthebibliography\thebibliography
\renewcommand\thebibliography[1]{
   \OLDthebibliography{#1}
   \setlength{\parskip}{0pt}
   \setlength{\itemsep}{0pt plus 0.3ex}
}

\begin{document}
\begin{titlepage}
\pubblock

\vfill
\Title{The GPD program at COMPASS}
\vfill
\Author{ Andrzej Sandacz\support\\
~~~~~~\\
(on behalf of the COMPASS Collaboration)}
\Address{\ncbj}
\vfill
\begin{Abstract}

The 160 GeV polarised muon beam available at CERN, with positive or
negative charge, makes COMPASS a unique place for GPD studies.
The first GPD related COMPASS results come from exclusive vector meson
production on transversely polarised protons and deuterons. The data were taken
in 2003-2010 with large solid-state polarised targets, 
although without detection of recoil
particles. Results on various transverse target spin dependent azimuthal
asymmetries are presented and their relations to GPDs are discussed.

The dedicated COMPASS GPD program started
in 2012 with commissioning of
a new long liquid hydrogen target and
new detectors such as
the large recoil proton detector
and the large-angle electromagnetic calorimeter. It was followed by a short
pilot 'DVCS run'. The performance of the setup and
first results on DVCS and exclusive $\pi ^0$ channels have been demonstrated.
The full data taking for the GPD program approved within COMPASS-II proposal
is planned for 2016 and 2017.

\end{Abstract}
\vfill
\begin{Presented}
Twelfth Conference on the Intersections of Particle and Nuclear Physics\\
~~~~~~~\\
Vail, Colorado, USA, May 19-24, 2015
\end{Presented}
\vfill
\end{titlepage}
\def\thefootnote{\fnsymbol{footnote}}
\setcounter{footnote}{0}

\section{Introduction}
General Parton Distributions (GPDs) \cite{AS_Mueller,AS_Ji,AS_Radyu} contain a wealth of information
on the partonic structure of the nucleon, which is the one of central problems
in hadron physics. In particular, GPDs allow a novel description of the nucleon as an extended object,
sometimes referred to as 3-dimensional 'nucleon tomo\-gra\-phy'~\cite{AS_Burkardt}. GPDs also allow access to such a fundamental
property of the nucleon as the orbital angular momentum of quarks~\cite{AS_Ji}.
For reviews of the GPDs see Refs \cite{AS_Goeke, AS_Diehl, AS_BelRad}.
The mapping of the nucleon GPDs requires
comprehensive experimental studies of hard processes such as Deeply Virtual Compton Scattering
(DVCS) and Hard Exclusive Meson
Production (HEMP) in a broad kinematic range.

\section{Exclusive $\rho ^0$ and $\omega $ muoproduction on transversely polarised protons}

The COMPASS collaboration has analysed exclusive vector meson production on 
polarised $^6$LiD 
(deuterons) and
NH$_3$ (protons) targets using the data from 2003-2010. Here the results for 
the data 
taken with polarised protons will be discussed. Although no recoil proton detector 
was included in the used experimental setup, which is a disadvantage for
measurements of exclusive processes, the analysis of these data allows us
to obtain 
first valuable results that are sensitive to 'elusive' GPDs $E$ and chiral-odd GPDs. 

\subsection{Experimental setup and event selection}

COMPASS is a fixed-target experiment situated at the high-intensity M2 beam-line of the
CERN SPS. A detailed description can be found in Ref.~\cite{Abbon:2007pq}.

The $\mu^+$ beam had a nominal momentum of 160 GeV/$c$ with a spread of 5\% and
a longitudinal polarisation of $P_{\ell}\approx-0.8$.  The data were taken at a
mean intensity of $3.5 \cdot 10^8\,\mu$/spill, for a spill length of about 10~s
every 40~s. A measurement of the trajectory and the momentum of each incoming
muon is performed upstream of the target. The momentum of the beam muon is
measured with a relative precision better than 1$\,\%$.

The beam traverses a solid-state ammonia (NH$_3$) target that provides
transversely polarised protons. The target is situated within a large aperture
magnet with a dipole holding field of 0.5~T. The 2.5~T solenoidal field is only
used when polarising the target material. A mixture of liquid $^3$He and $^4$He
is used to cool the target to 50~mK. The ammonia is contained in three cylindrical
target cells with a diameter of 4~cm, placed one after another along the
beam. The central cell is 60~cm long and the two outer ones are 30~cm long, with
5 cm space between cells. The spin directions in neighbouring cells are
opposite. Systematic effects due to acceptance are
reduced by reversing the spin directions on a weekly basis.

The COMPASS two-stage spectrometer is designed to reconstruct scattered 
muons and
produced hadrons in wide momentum and angular ranges. Each stage has a
dipole magnet with tracking detectors before and after the magnet, hadron and
electromagnetic calorimeters and muon identification. Identification of charged
tracks with a RICH detector in the first stage is not used for the results
presented here.
%

To determine the transverse target spin asymmetries for exclusive $\rho ^0$
production the data taken in 2007 and 2010 with polarised protons are used.
The details of selection of the sample are given in Ref.~\cite{Adolph:2012np}.
The
essential steps of event selection and asymmetry extraction are summarised in
the following. The considered events are characterised by an incoming and a
scattered muon and two oppositely charged hadrons, $h^+h^-$, with all four
tracks associated to a common vertex in the polarised target. In order to select
events in the deep inelastic scattering regime and suppress radiative
corrections, the following cuts are used: $Q^2 > $ 1 (GeV/$c$)$^2$, 0.003
$<x_{\mathrm{Bj}}<$ 0.35, $W>$ 5 GeV and 0.1$<y<$ 0.9. The production of $\rho ^0$
mesons is selected in the two-hadron invariant mass range 0.5 GeV/$c^2$ $<
M_{\pi ^+\pi ^-} <$ 1.1 GeV/$c^2$, where for each hadron the pion mass
hypothesis is assigned. This cut is optimised towards high yield and purity of
$\rho ^0$ production, as compared to non-resonant $\pi^+\pi^-$ production.  The
measurements are performed without detection of the recoiling proton in the
final state. Exclusive events are selected by choosing a range in missing
energy,
\begin{equation}
E_{\text{miss}} = \frac{(p+q-v)^2 - p^2}{2 M_{\text{p}}}.
\end{equation}
The four-momenta of proton, photon, and meson, are denoted by $p$, $q$, and $v$
respectively, and $M_{\text{p}}$ is the proton mass. Although for exclusive events $E_{\text{miss}}$ $\approx$ 0 holds,
the finite experimental resolution is taken into account by selecting events in
the 'signal' range $ |E_{\text{miss}} | <$ 2.5 GeV, which corresponds to $0 \pm 2\sigma$
where $\sigma$ is the width of the Gaussian signal peak. Non-exclusive
background is further suppressed by cuts on the squared transverse momentum of the
vector meson with respect to the virtual-photon direction, $p_\mathrm{T}^2 <$ 0.5
(GeV/$c$)$^2$, the energy of the $\rho ^0$ in the laboratory system, $E_{\rho
  ^0} >$ 15 GeV, and the photon virtuality, $Q^2<$ 10 (GeV/$c$)$^2$. An
additional cut $p_\mathrm{T}^2>0.05$ (GeV/$c$)$^2$ is used to reduce the 
contribution of events from coherent production on the target nuclei.
The average
values of the kinematic variables are $\langle Q^2 \rangle = 2.15 $
(GeV/$c$)$^2$, $\langle x_{\mathrm{Bj}} \rangle = 0.039$, $\langle y \rangle = 0.24$,
$\langle W \rangle = 8.13$ GeV, and $\langle p_\mathrm{T}^2 \rangle = 0.18$
(GeV/$c$)$^2$.
Details on the method to correct for the contribution of the semi-inclusive background
in the sample and on the extraction of asymmetries are given in Ref.~\cite{Adolph:2014pl}.

The analysis of exclusive $\omega $ production was performed using the data taken 
in 2010 with transversely polarised protons.  An event to be accepted for
further analysis was required to have an incident muon track, a scattered
muon track, exactly two additional tracks of oppositely charged
hadrons, all associated to a vertex in the polarised target
material, and a single $\pi ^0$ meson that is reconstructed using its two decay photons detected
in the electromagnetic calorimeters. The photon clusters in the calorimeters have to be time-correlated with the beam track.

Most of the selections of events from exclusive $\omega $ production follow those for $\rho ^0$ analysis, except these listed 
in the following. In order to select $\pi ^0$ meson, a cut on the accepted range of invariant mass of 
two photons, $M_{\gamma \gamma}$, was applied, which took into account the energy-dependent resolution of photon energy measurements. The $\omega $ resonance was selected
by the cut on the invariant mass of $\pi ^+ \pi ^- \pi ^0$ system, 
$| M_{\pi \pi \pi} - M_{\omega}^{\mathrm{PDG}} | < 70\,\mathrm{MeV}/c^2$, where
$M_{\omega}^{\mathrm{PDG}} = 782.65\,$MeV$/c^2$. Due to an additional contribution
of calorimetric measurements to the experimental resolution, the exclusivity selection
was changed to $ |E_{\text{miss}} | <$ 3.0 GeV. In the analysis of $\omega $
meson the signal and background asymmetries were extracted simultaneously using the unbinned maximum likelihood method. The average values of the kinematic variables are
$\langle Q^2 \rangle = 2.2 $
(GeV/$c$)$^2$, $\langle x_{\mathrm{Bj}} \rangle = 0.049$, $\langle y \rangle = 0.18$,
$\langle W \rangle = 7.1$ GeV, and $\langle p_\mathrm{T}^2 \rangle = 0.17$
(GeV/$c$)$^2$.
 
\subsection{Results and discussion}

\begin{wrapfigure}[21]{R}{80mm}
\hspace{-18pt}
\includegraphics[width=80mm]{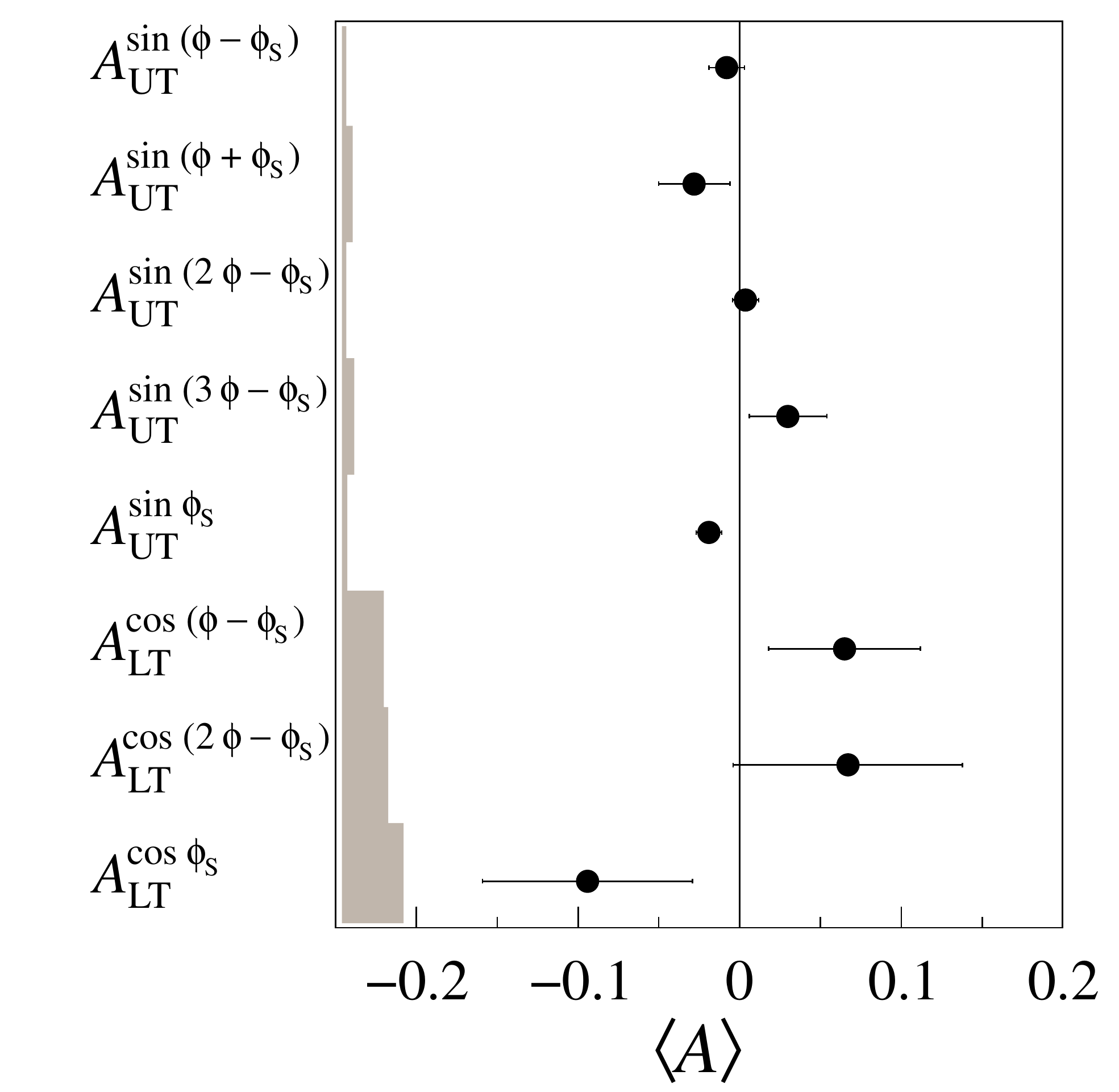}
\caption{\footnotesize Mean values $\langle A \rangle$ for all transverse target 
asymmetries for exclusive $\rho ^0$ productions. The errors bars (left bands)
represent the statistical (systematic) uncertainties.}
\label{fig:rho}
\end{wrapfigure}

For a transversely polarised target five single ($\rm{UT}$) and three double ($\rm{LT}$) spin asymmetries can be defined. 
These are $ A_{\rm{UT}}^{\sin \left( \phi - \phi_{s} \right)}$,
$A_{\rm{UT}}^{\sin \left( \phi + \phi_{s} \right)}$,
$A_{\rm{UT}}^{\sin \left(3\phi - \phi_{s} \right)}$,
$A_{\rm{UT}}^{\sin \phi_{s}}$,
$A_{\rm{UT}}^{\sin \left(2\phi - \phi_{s} \right)}$,
$A_{\rm{LT}}^{\cos \left( \phi - \phi_{s} \right)}$,
$A_{\rm{LT}}^{\cos \phi_{s}}$, 
$A_{\rm{LT}}^{\cos \left(2\phi - \phi_{s} \right)}$.
Each asymmetry is related to a modulation of the cross section as a function of \aphi and \aphis angles, which is indicated by the superscript. 
The angle \aphi is the azimuthal angle between the lepton plane, given by the momenta of the incoming and the scattered leptons, and the hadron plane, given by the momenta of the virtual photon and the meson. The angle \aphis is the azimuthal angle between the lepton plane and the spin direction of the target nucleon.
Average values of measured asymmetries for exclusive $\rho ^0$ production are shown 
in Fig.~\ref{fig:rho}. The asymmetry
\AUTphiS was found to be $-0.019 \pm 0.008(stat.) \pm 0.003(syst.)$. All other
asymmetries were also found to be of small magnitude but consistent with zero
within experimental uncertainties. The asymmetries measured as functions of
$Q^2$, $x_{\mathrm{Bj}}$ and $p_\mathrm{T}^2$ can be found in Ref.~\cite{Adolph:2014pl}, as well as their 
comparison to the predictions of the GPD-model by Goloskokov and Kroll \cite{Goloskokov:2014epj}.
 
The measured asymmetries are proportional to bilinear combinations of the helicity 
amplitudes $\mathcal{M}$ for the photoproduction subprocess, 
$A \propto \sum \mathcal{M}_{i'm',im}^{*}\mathcal{M}_{i'm',jn}$,
where the helicity of the virtual photon is denoted by $i, j = -1, 0, +1$ and the helicity of the initial-state proton by $m, n = -\frac{1}{2}, +\frac{1}{2}$. The sum runs over all combinations of spins given by the spin of the meson $i' = -1, 0, +1$ and the spin of the final-state proton $m' = -\frac{1}{2}, +\frac{1}{2}$. In the following the helicities  will be labelled by only their sign or zero.

For an interpretation of results in the framework of the model, the following asymmetries are particularly interesting, for which the dependence on the helicity amplitudes reads
\begin{align}
\sigma_{0} A_{\mathrm{UT}}^{\sin \left( \phi - \phi_{s} \right)\text{\phantom{$3$}}}& =
-2 \rm{Im} \left[ \epsilon \mathcal{M}_{0-,0+}^{*} \mathcal{M}_{0+,0+} + \mathcal{M}_{+-,++}^{*} \mathcal{M}_{++,++} + \tfrac{1}{2} \mathcal{M}_{0-,++}^{*} \mathcal{M}_{0+,++} \right]~, \nonumber \\[0pt]
\sigma_{0} A_{\mathrm{UT}}^{\sin \left(2\phi - \phi_{s} \right)} & =
-\phantom{2} \rm{Im} \left[ \mathcal{M}_{0+,++}^{*} \mathcal{M}_{0-,0+} \right]~, \nonumber \\[0pt]
\sigma_{0} A_{\mathrm{UT}}^{\sin \phi_{s}\text{\phantom{$\left( 3\phi + \right) $}}}& =
-\phantom{2} \rm{Im} \left[ \mathcal{M}_{0-,++}^{*} \mathcal{M}_{0+,0+} - \mathcal{M}_{0+,++}^{*} \mathcal{M}_{0-,0+} \right]~.
\end{align}

Here $\sigma_{0}$ is the total unpolarised cross section, given by the sum of cross sections for longitudinally, $\sigma_{L}$, and transversely, $\sigma_{T}$, polarised virtual photons,
\begin{equation}
\sigma_{0} = \frac{1}{2}\left( \sigma_{++}^{++} + \sigma_{++}^{--} \right) + \epsilon \sigma_{00}^{++} = \sigma_{L} + \epsilon \sigma_{T} , 
\label{eq:theory:05}
\end{equation}
and $\epsilon $ is the virtual photon polarisation parameter.

The dominant $\gamma^*_L \rightarrow \rho^0_L$ transitions are
described by helicity amplitudes $\mathcal{M}_{0+,0+}$ and
$\mathcal{M}_{0-,0+}$, which are related to chiral-even GPDs $H$ and
$E$, respectively. The subscripts $L$ and $T$ denote the photon and
meson helicities $0$ and $\pm1$, respectively. The suppressed
$\gamma^*_T \rightarrow \rho^0_T$ transitions are described by the
helicity amplitudes $\mathcal{M}_{++,++}$ and $\mathcal{M}_{+-,++}$,
which are likewise related to $H$ and $E$. These GPDs are used
since several years to describe DVCS and vector meson exclusive production data.
By the recent inclusion of
transverse, {\it i.e.} chiral-odd GPDs, it became possible to also describe
$\gamma^*_T \rightarrow \rho^0_L$ transitions~\cite{Goloskokov:2014epj}. 
In their description
appear the amplitudes $\mathcal{M}_{0-,++}$ related to chiral-odd GPDs
${H}_T$  and $\mathcal{M}_{0+,++}$
related to chiral-odd GPDs $\overline {E}_T$. 

The small value of $A_{\mathrm{UT}}^{\sin \left( \phi - \phi_{s} \right)}$ asymmetry 
for $\rho ^0$ is explained as an approximate cancellation of contributions from GPDs $E^u$and $E^d$ for valence quarks $u$ and $d$, respectively, which have opposite signs but similar absolute values.
The $A_{\mathrm{UT}}^{\sin \phi_{s}}$ asymmetry represents an imaginary part of two bilinear products of helicity amplitudes. The first product is related to GPDs $H$ and $H_{T}$, while the second one is related to GPDs $E$ and $\overline E_{T}$. 
The asymmetry \AUTphi represents the same combination of
GPDs $E$ and $\overline {E}_T$ as the second term in \AUTphiS. The
observation of a vanishing value for \AUTphi implies that the
non-vanishing value of \AUTphiS constitutes the first experimental
evidence from hard exclusive $\rho^0$ leptoproduction for the
existence of transverse GPDs $H_T$. 

\begin{figure}[htb]
\centering
\vspace*{0mm} 
\raisebox{-0.5\height}{\includegraphics[width=75mm]{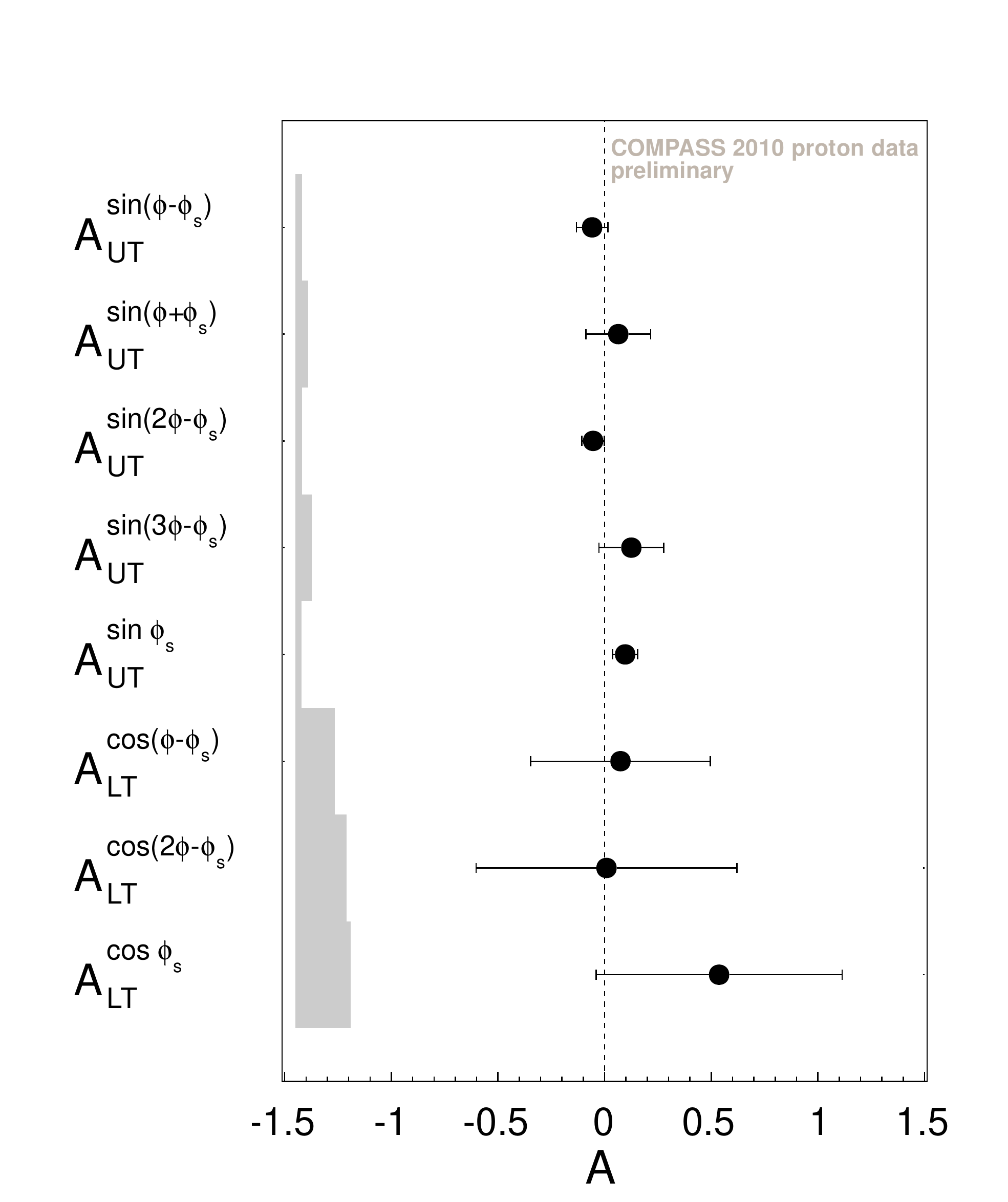}}
\raisebox{-0.5\height}{\includegraphics[width=75mm]{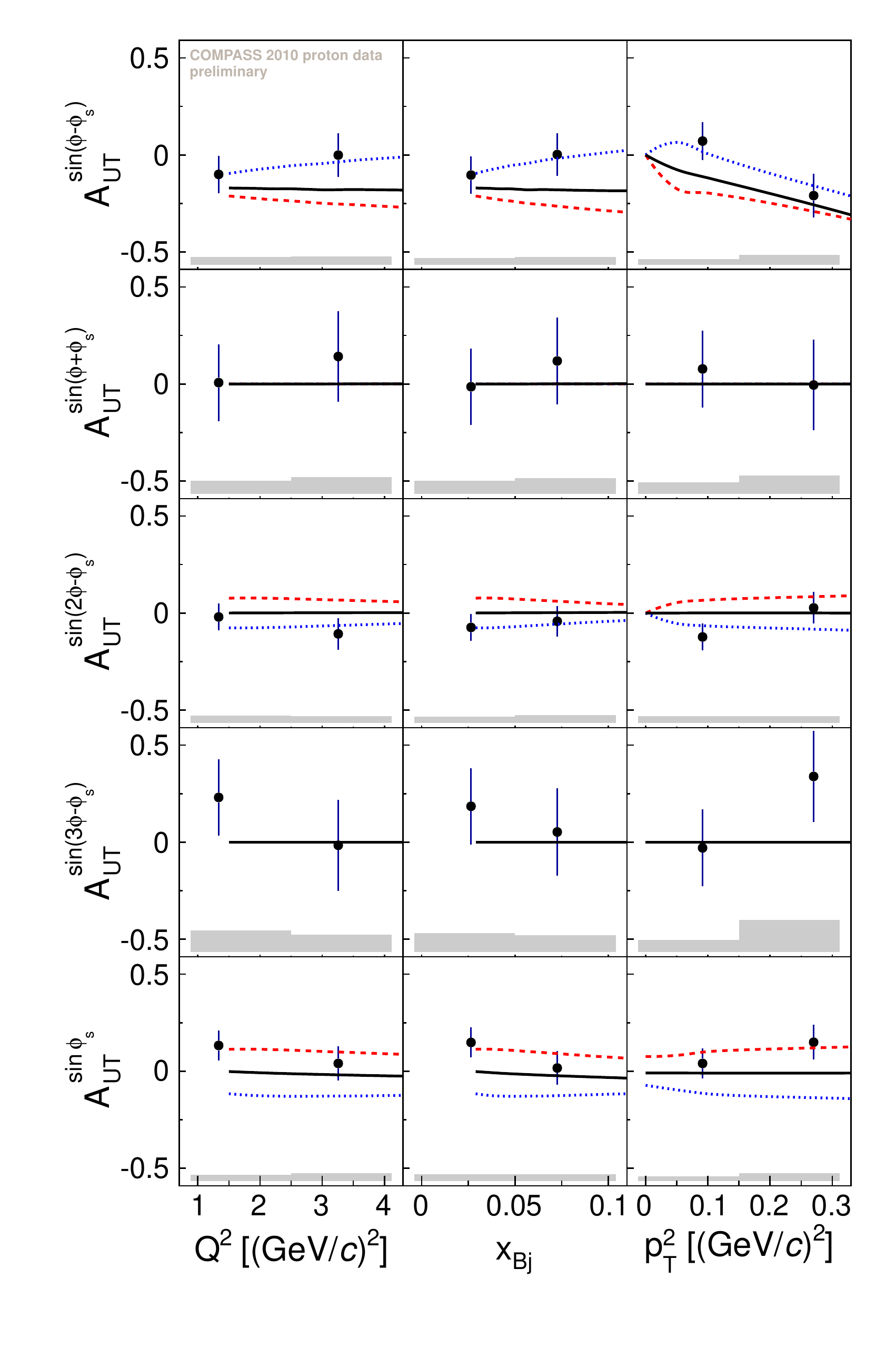}}
\caption{\footnotesize Left: Mean values $A$ for all transverse target
asymmetries for exclusive $\omega $ production. The errors bars (left bands)
represent the statistical (systematic) uncertainties. Right: Single transverse target
spin asymmetries for exclusive $\omega $ production. The curves show the predictions of
the GPD model of Ref.~\cite{Goloskokov:2014epj}. The dashed red and dotted blue lines
represent the predictions with the positive and negative $\pi \omega $ form factors,
respectively, while the solid black lines represent the predictions without the pion pole exchange.}
\label{fig:omega}
\end{figure}

The preliminary result on average values of all eight transverse target asymmetries 
for exclusive $\omega $ production are shown
in Fig.~\ref{fig:omega} (left), while the single spin asymmetries as functions of
$Q^2$, $x_{\mathrm{Bj}}$ and $p_\mathrm{T}^2$ are presented in Fig.~\ref{fig:omega} (right) and are 
compared
to the predictions of Goloskokov-Kroll model \cite{ref_10d}. In this version
of the model the authors have added a contribution from the exchange of $\pi ^0$
pole. It turned out that this is an important contribution needed to reproduce
HERMES results on SDMEs for exclusive
electroproduction of $\omega $ mesons \cite{ref_10c}. Still, the SDME data do not allow one to distinguish
the sign of the $\pi \omega$ transition form factor. Certain azimuthal asymmetries
for $\omega $ production are sensitive to the pion pole contribution and
hence in principle could allow one to determine the sign of the form factor. The effect of the
pion pole decreases with the energy of the virtual photon nucleon system, $W$,
but still it could be measurable at
COMPASS energies. 
The authors
have provided three sets of predictions: one without the pion pole contribution
and two others with the contributions that differ by the sign of $\pi \omega $
form factor. While the negative sign is preferred by the results on 
$A_{\rm{UT}}^{\sin \left( \phi - \phi_{s} \right)}$ and
$A_{\rm{UT}}^{\sin \left(2\phi - \phi_{s} \right)}$, the results on 
$A_{\rm{UT}}^{\sin \phi_{s}}$ asymmetry prefer the positive sign.
At present this discrepancy is not resolved.


\section{The GPD program of COMPASS-II}
The GPD part of COMPASS-II proposal~\cite{proposal} is devoted to measurements of both DVCS
and HEMP with polarised $\mu ^+$ and $\mu ^-$ beams and a liquid hydrogen
target. The new detectors, the 4 m-long
recoil proton detector CAMERA and (a central part of) the new large-angle 
electromagnetic calorimeter ECAL0, 
which are essential for measurements of exclusive processes, were constructed and incorporated into the COMPASS setup in 2012. The commissioning of these new 
detectors was done in 2012 and was followed by
a short DVCS pilot run. The dedicated data taking for the GPD program, with
the complete calorimeter ECAL0,
is foreseen in 2016-2017 for a total period of 280 days.

The recoil proton detection is based on the ToF measurement between two barrels 
of 24 scintillator slats read out at both ends. By installing the CAMERA around
the new 2.5 m long LH$_2$ target COMPASS has been
converted into a facility measuring exclusive reactions within a kinematic
domain from $x_{\mathrm{Bj}} \sim 0.01$ to $\sim 0.1$, which cannot be explored
at any other existing or planned facility in the near future.
The new calorimeter ECAL0 is being constructed.
Compared to the existing electromagnetic calorimeters, it will increase the accessible
kinematic domain for DVCS and exclusive $\pi ^0$ production towards
higher $x_{\mathrm{Bj}}$, and therefore it will
provide an overlap with HERMES and JLAB experiments. ECAL0 will also improve the hermeticity
for detection of exclusive events and will contribute to reduce background to single-photon
production that originates from $\pi ^0$ and other decays.

An efficient selection of exclusive events,
and suppression of the background
was possible by using the combined information from the forward COMPASS 
detectors and the CAMERA. As an example a result for exclusive single
photon production, obtained from the 2012 DVCS pilot run, is presented in 
the following.

\begin{figure}[htb]
\centering
\includegraphics[height=3.0in]{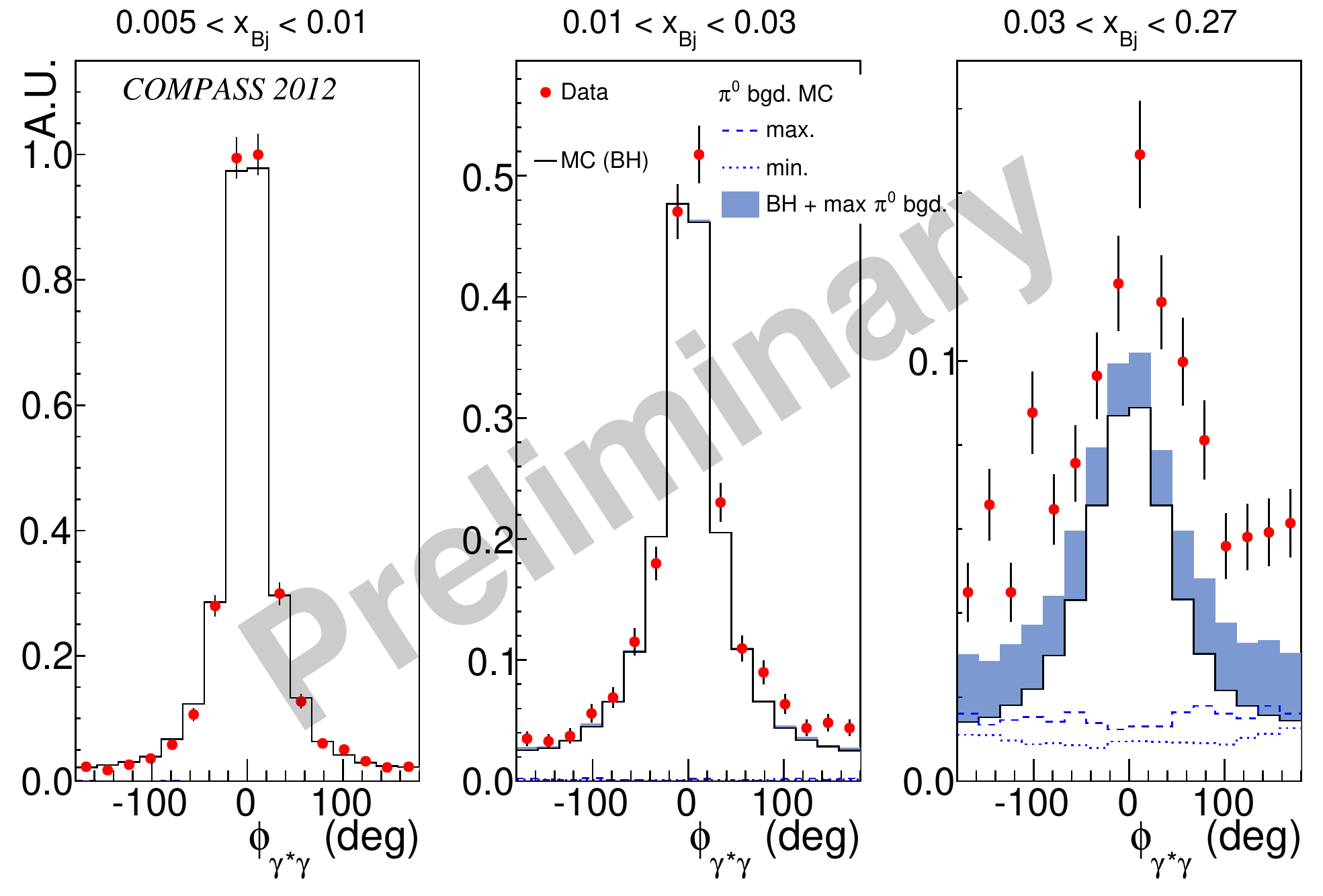}
\caption{\footnotesize The exclusive single photon events obtained from the 
2012 sample as a function of $\phi _{\gamma *\gamma }$ 
($\equiv \phi$)
compared to the MC estimates of BH contribution (solid line) and the $\pi ^0$ contamination (dotted and dashed lines). See text for more details.
}
\label{fig:dvcs}
\end{figure}

A way to identify the observed process, $\mu + p \rightarrow \mu ' + \gamma +p'$, to which both the DVCS and Bethe-Heitler (BH) processes
contribute, is to look at the angle $\phi $ between the leptonic plane and 
the $\gamma ^* \gamma$ 
plane. The observed distributions, after applying all cuts and selections
and for $Q^2 > 1 \: ({\rm GeV}/c)^2$, are displayed in Fig.~\ref{fig:dvcs} and compared to the
predictions from the Monte Carlo (MC) simulations for the BH event yield.
The MC is normalised to the data in the small $x_{\mathrm{Bj}}$ bin,
where only BH contribution is expected. 
COMPASS offers the advantage to provide various kinematic domains where either  BH or DVCS
dominates. The collection of almost pure BH events at small $x_{\mathrm{Bj}}$ allows one to get an excellent reference yield and
to control accurately the global efficiency of the apparatus.
In contrast, the collection of a DVCS sample at larger $x_{\mathrm{Bj}}$ will allow the measurement of the
$x_{\mathrm{Bj}}$ dependence of the $t$-slope of the cross section, which is related to the tomographic partonic image of the nucleon.
In the intermediate domain, the DVCS contribution will be boosted by the BH 
process through the interference term that allows us to investigate DVCS at the 
amplitude level.
The Bethe-Heitler contribution shows a characteristic peak at $\phi \simeq 0$. 
 
The exclusive photon production sample is polluted by $\pi ^0$ contribution
when only one photon of the $\pi ^0$ decay is observed. In this case, one of
the two photons carries most of the energy of the parent meson and the second 
photon us either absorbed or emitted outside of the ECALs acceptance. We call
this a "non-visible" $\pi ^0$ background, which affects mostly large
$x_{\mathrm{Bj}}$ events. It comes both from exclusive and semi-inclusive 
(SIDIS) $\pi ^0$
productions. The first was studied using HEPGEN MC for hard exclusive $\pi ^0$
production, while the second one by using LEPTO MC. The MC simulations have
been normalised to the real data by comparing reconstructed $\pi ^0$ in the 
real data with all reconstructed $\pi ^0$ from each of the two MC generators
separately. Presently we use a conservative approach and take the SIDIS $\pi ^0$ (LEPTO)
sample
to estimate an upper limit and the exclusive $\pi ^0$ (HEPGEN) sample for
a lower bound on invisible $\pi ^0$ contribution to the single photon sample.
These two estimates are shown in Fig.~\ref{fig:dvcs} as dotted (exclusive $\pi ^0$) and dashed (SIDIS $\pi ^0$) histograms. The sum of BH contribution (continuous line)
and the estimate of the maximal $\pi ^0$ background is shown as a shaded
(blue) histogram (right panel). An excess of the data above this histogram is
interpreted as an indication of the DVCS events in the sample and a proof of 
feasibility to measure DVCS with the present setup.

The projected accuracies for measurements of various observables within the
GPD program are presented in the COMPASS-II proposal~\cite{proposal}. 
Investigation of GPDs with DVCS and HEMP on unpolarised protons using the data
from 2016-2017 will allow to determine
the $x_{\mathrm{Bj}}$-dependence for $t$-slopes of the differential cross sections. That is related
to the transverse distribution of partons and the 'nucleon tomography'. Measurements of
the beam charge and spin sum and difference of single-$\gamma $ cross sections will give
access
to the real and imaginary parts of the DVCS amplitude, and will allow to further constrain
GPDs $H$. Studies of exclusive production of vector mesons ($\rho $, $\omega $, $\phi $)
will lead to the quark flavour and gluon separation for GPDs $H$, while that of exclusive
$\pi ^0$ production will provide constrains on the GPD $\widetilde{E}$ and on chiral-odd GPDs.

The main goal of future (after 2018) measurements with transversely polarised target is to constrain
GPDs $E$, which are related to the orbital momentum of partons, and also to investigate
the role of chiral-odd GPDs in exclusive meson production.

\end{document}